\documentclass[aps,prb,reprint,floatfix,superscriptaddress,showpacs]{revtex4-1}
\pdfoutput=1
\usepackage{epsfig}
\usepackage{amsmath}
\usepackage{array}
\usepackage{hyphenat}
\usepackage{hyperref}
\usepackage{comment}
\usepackage{placeins}
\usepackage{dsfont}
\usepackage{graphicx}
\usepackage{pgfplotstable, booktabs}
\usepackage{color}
\hypersetup{
  colorlinks,
  citecolor=magenta,
  linkcolor=blue,
  urlcolor=blue}

\usepackage[justification=raggedright]{caption}	
\usepackage[colorinlistoftodos]{todonotes}

\begin{document}
\title{R\'{e}nyi entanglement entropy of critical SU($N$) spin chains}
\author{Jonathan D'Emidio}
\affiliation{Department of Physics \& Astronomy, University of Kentucky, Lexington, KY 40506-0055}
\author{Matthew S. Block}
\affiliation{Department of Physics \& Astronomy, California State University, Sacramento, CA 95819}
\author{Ribhu K. Kaul}
\affiliation{Department of Physics \& Astronomy, University of Kentucky, Lexington, KY 40506-0055}
\begin{abstract}

We present a study of the scaling behavior of the R\'{e}nyi entanglement entropy (REE) in SU($N$) spin chain Hamiltonians, in which all the spins transform under the fundamental representation. These SU($N$) spin chains are known to be quantum critical and described by a well known Wess-Zumino-Witten (WZW) non-linear sigma model in the continuum limit.  Numerical results from our lattice Hamiltonian are obtained using  stochastic series expansion (SSE) quantum Monte Carlo for both closed and open boundary conditions. As expected for this 1D critical system, the REE shows a logarithmic dependence on the subsystem size with a prefector given by the central charge of the SU($N$) WZW model.  We study in detail the sub-leading oscillatory terms in the REE under both periodic and open boundaries.  Each oscillatory term is associated with a WZW field and decays as a power law with an exponent proportional to the scaling dimension of the corresponding field.  We find that the use of periodic boundaries (where oscillations are less prominent) allows for a better estimate of the central charge, while using open boundaries allows for a better estimate of the scaling dimensions.  For completeness we also present numerical data on the thermal R\'{e}nyi entropy, which equally allows for extraction of the central charge.

\end{abstract}


\setlength{\belowcaptionskip}{-8pt}
\maketitle
\section{Introduction}  

Entanglement is a fascinating aspect of many-body quantum systems.~\cite{Amico2008:Entanglement}  It describes how a many-body wave function cannot in general be written as a tensor product of individual single-particle wave functions.  The degree to which a wave function fails to be written as a product state of two subsystem wave functions can be quantified in terms of the entanglement entropy (EE) between these two subsystems. In one dimension, gapped quantum systems have an EE that stays constant as the size of the subsystem is increased. This is consistent with the so-called ``area law,'' which states that the EE grows with the area of the boundary of the subsystem.\cite{Eisert2010:AreaLaws} However in gapless conformally invariant systems, the EE violates the area law and exhibits a logarithmic divergence with a prefactor given by the central charge. \cite{Holzhey1994:GeometricEntropyCFT,Calabrese2004:EEandQFT,Korepin2004:thermal} The logarithmic growth of entanglement has been observed in a number of quantum critical lattice models whose scaling limit is described by a CFT.  

The entanglement entropy also has sub-leading terms that contain universal information beyond the central charge. Namely, one can find power law dependencies on the subsystem size with exponents given by the scaling dimensions of various conformal field theory (CFT) operators.\cite{Cardy2010:Unusual}  Indeed, such universal contributions have been observed both numerically\cite{laflorencie2006boundary,Calabrese2010:ParityEffects} and analytically.\cite{Fagotti2011:UniversalParity}  These sub-leading terms in the entanglement entropy can be used to characterize the operator content of the underlying CFT of various lattice models.\cite{Xavier2012:FiniteSizeCorrectionsEE, Xavier2011:ParitySpinS,Dalmonte2011:EstimatingOrder}  

In this work we consider the R\'{e}nyi entanglement entropy (defined in Sec. \ref{sec:data})  in the context of SU($N$) Heisenberg antiferromagnetic spin chains. We are able to completely characterize the underlying CFT for this model based on the central charge from the leading log term, as well as the operator content contained in the sub-leading oscillatory terms, which decay as power laws.   We have also measured the R\'{e}nyi entropy of a subsystem as a function of temperature in order to gain perspective on finite temperature effects and as a secondary means of extracting the central charge.

\section{The lattice Hamiltonian}

We consider the following Hamiltonian with the spin on each lattice site transforming under the fundamental representation of the SU($N$) algebra.  
{\allowdisplaybreaks
\begin{equation}
		H_{\Pi} = J\sum_{<ij>}\sum^{N-1}_{\alpha ,\beta =0} |\beta_{i}\alpha_{j}\rangle\langle\alpha_{i}\beta_{j}|
\label{eq:PermutationHamiltonian}
\end{equation}}
This model\cite{Sutherland1975:MulticompModel} consists of a sum of bond operators that permute nearest neighbor spins (colors) which are labeled by numbers 0 through $N-1$.  $H_\Pi$ reduces to the spin-$1/2$ Heisenberg model when $N=2$, and provides a natural extension to larger $N$. We will consider the case when $J$ is positive (antiferromagnetic) and spins tend to anti-align with one another. Since it takes $N$ lattice sites to form an SU($N$) singlet, we expect and find that the ground state consists of equal numbers of each color and is non-degenerate if the chain is an integer multiple of $N$.  Finally, and most importantly for the work that we present here, the ground state is described by the SU($N$)$_{1}$ WZW model with central charge $c=N-1$\cite{Affleck1986:CriticalExpSUN,Affleck1988:CriticalSUN} and $N-1$ primary fields with scaling dimensions $\Delta_{a}=a-a^{2}/N$ where $a \in [1,N-1]$.\cite{Bouwknegt1999:ExclusionCFTWZW,Assaraf1999:MetalInsulator}

Models with SU($N$) symmetry are of both theoretical and experimental interest, since it has been shown that ultra cold atoms in optical traps can give rise to this symmetry.\cite{Gorshkov2010:TwoOrbitalSUN}  In fact, the model we consider here can be obtained from the SU($N$) Hubbard model at 1/$N$ filling in the limit of large on-site repulsion (see [\onlinecite{Assaraf1999:MetalInsulator},\onlinecite{Manmana2011:SUNmagnetismUCA}] for a numerical study of the Mott transition).

Entanglement in this class of models has previously been studied using DMRG for $N\leq 4$,\cite{Fuhringer2008:DMRGSUN} though the universal sub-leading oscillations were not present in the von Neumann entanglement entropy under closed boundary conditions.  Other studies\cite{Frischmuth1999:ThermoSU4,Messio2012:entropyDep} found oscillations in the spin-spin correlation function for different values of $N$, verifying that the periodicity is given by 2$\pi/N$.  In fact, one can make a precise connection between entanglement and spin correlations in one dimension.\cite{Song2010:EntanglementFluctuations}

Here we will study in detail the scaling form of these oscillations which are induced in the R\'{e}nyi entanglement entropy under both open and closed boundaries. We will demonstrate quantitatively that the decay of these oscillations contain interesting information about the scaling dimensions of operators in the WZW field theory.

\begin{figure}[!t]
\centerline{\includegraphics[angle=0,width=0.95\columnwidth]{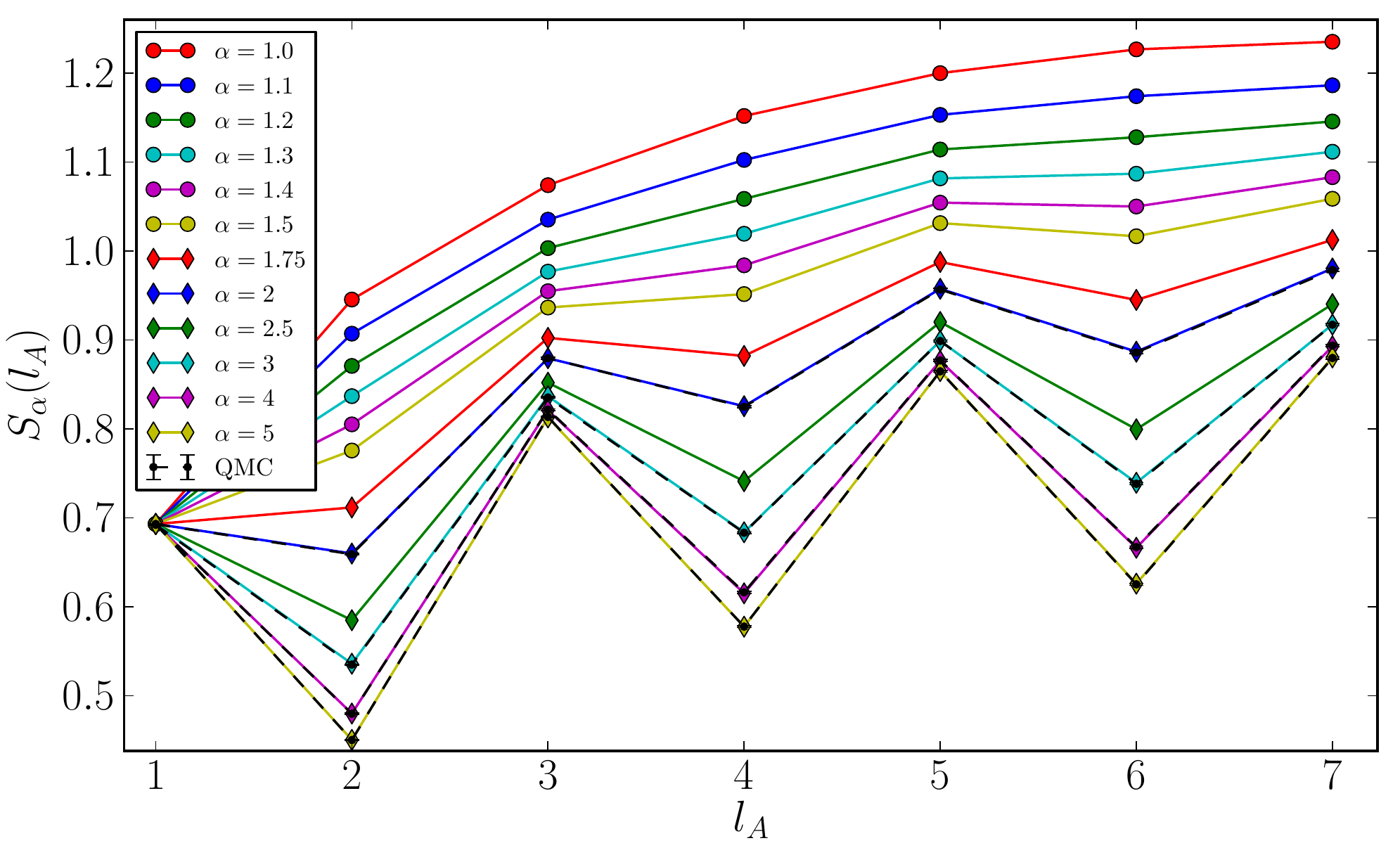}}
\caption{Comparison of QMC and exact diagonalization for the ground state REE of a 14 site chain with $N$=2 under periodic boundary conditions. This is shown for different values of the R\'{e}nyi index $\alpha$, as indicated in the legend.  Colored points are data from exact diagonalization and black circles connected by dashed lines are obtained from QMC for $\alpha=2,3,4,5$. Note that under periodic boundaries, oscillations in the REE appear when $\alpha>1$ (see the end of Appendix \ref{app_replica}).}
\label{fig:EDsu2}
\end{figure}

\section{Numerical results} 
\label{sec:data}

We begin by first defining the R\'{e}nyi entanglement entropy (REE).  Take a one-dimensional system of length $L$ and partition it into two segments of length $l_{A}$ and $l_{B}=L-l_{A}$.  Construct the density matrix for the entire system ($\rho$), and compute the reduced density matrix ($\rho_{A}=\mathrm{Tr}_{B}\rho$) by tracing over the degrees of freedom in $l_{B}$.  The R\'{e}nyi entanglement entropy is then given by
\begin{equation}
		S_{\alpha}(\rho_{A})= \frac{1}{1-\alpha}\log(\text{Tr}\{ \rho^{\alpha}_{A} \}),
\label{eq:REE}
\end{equation}
where one obtains the von Neumann entanglement entropy $S(l_{A})=-\textrm{Tr}\{\rho_{A}\log(\rho_{A})\}$ by taking the limit $\alpha\to1$.  Appendix \ref{app_replica} reviews the extended ensemble approach introduced in [\onlinecite{Humeniuk2012:ExtendedEnsemble}] that allows for efficient measurements of the REE with QMC.  In Fig. \ref{fig:EDsu2} we compare our QMC results versus exact diagonalization for an SU(2) chain of length $L=14$ under periodic boundaries for several values of the R\'{e}nyi index ($\alpha$).    The QMC delivers exact results within controllable statistical error bars.  Similar agreement is found for $N>2$ (not shown).

\subsection{REE periodic boundaries} 
\label{subsec:REEclosedBC}

Inspired by previous work on the XXZ spin chain we fit our numerical data to the following scaling form\cite{Calabrese2010:ParityEffects}
{\allowdisplaybreaks
\begin{equation}
		S_{\alpha}(l_{A}) =S^{\mathrm{log}}_{\alpha}(l_{A})+S^{\mathrm{osc}}_{\alpha}(l_{A})+\tilde{c}_\alpha,
\label{eq:GenScalingForm}
\end{equation}}%
where
{\allowdisplaybreaks
\begin{equation}
		S^{\mathrm{log}}_{\alpha}(l_{A}) =\frac{c}{6 \eta}\left(1+\frac{1}{\alpha}\right)\log\left[\left(\frac{\eta L}{\pi}\sin\left(\frac{\pi l_{A}}{L}\right)\right)\right]
\label{eq:Slog}
\end{equation}}%
and 
{\allowdisplaybreaks
\begin{equation}
		S^{\mathrm{osc}}_{\alpha}(l_{A}) =F_{\alpha}(l_{A}/L)\cos(2k_Fl_{A})\left|\frac{2\eta L}{\pi}\sin(\pi l_{A}/L)\right|^{-\frac{2\Delta_{1}}{\eta \alpha}},
\label{eq:Sosc}
\end{equation}}%
where $\eta=1,2$ is for periodic and open boundary conditions, respectively.  The universal parameters are the central charge $c$, the Fermi momentum $k_{F}$, and the scaling dimension $\Delta_{1}$.  $F_{\alpha}(l_{A}/L)$ is a universal scaling function (into which a factor of $|\sin(k_{F})|^{\frac{2\Delta_{1}}{\eta \alpha}}$ has been absorbed) and $\tilde{c}$ is a non-universal constant.  In the present study we find that approximating $F_{\alpha}(l_{A}/L)$ to be a constant allows for a sufficiently accurate extraction of the parameters of interest.  In the rest of this paper we take $k_{F}=\pi/N,$ which can clearly be seen in the data.  All of our simulations are performed with $\beta J=N L$, which ensures that finite temperature effects are negligible.

\begin{figure}
\centerline{\includegraphics[angle=0,width=1.0\columnwidth]{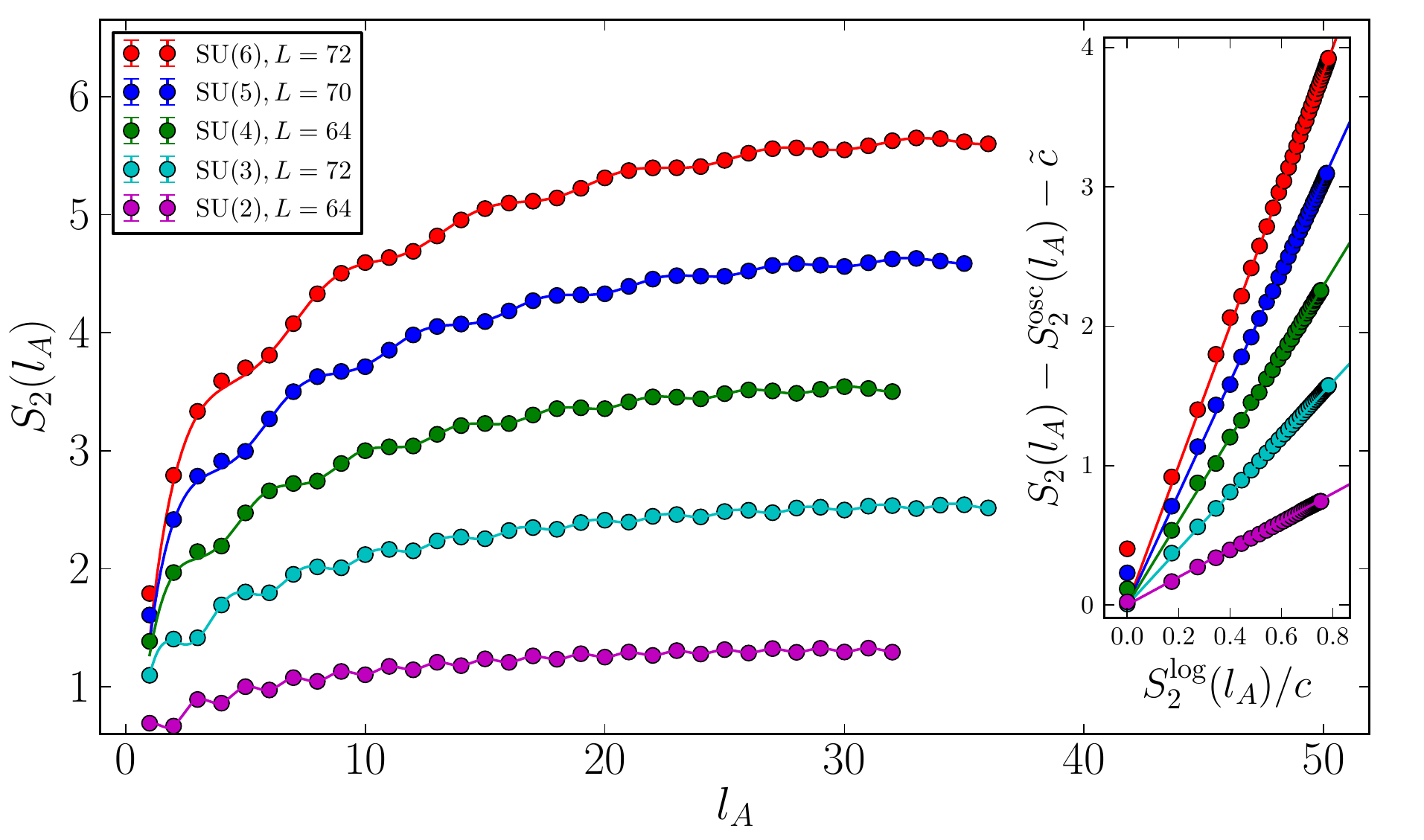}}
\caption{REE as a function of the subsystem size ($l_{A}$) with periodic boundary conditions.  We set the R\'{e}nyi index $\alpha = 2$ and the total chain lengths are integer multiples of $N$.  Oscillations have Fermi momentum $k_{F}=\pi / N$. Solid lines are best fits to the CFT scaling form.  In the right inset we have subtracted the oscillatory and constant pieces of the REE, and plotted it against $S_{\alpha}^{\mathrm{log}}(l_{A})/c$.  This is plotted on top of lines with slope $N-1$.  The best fit values for the central charges are given in Table \ref{tab:central}.}
\label{fig:REE}
\end{figure}

Fig. \ref{fig:REE} shows our data for the REE with periodic boundary condition for $2\leq N \leq 6$. When $N=2$, there is just one primary field with conformal weight $\Delta_{1}$.  In the case of higher $N$, there are primary fields with less relevant scaling dimensions that contribute in addition to the oscillatory behavior.  In the periodic case, oscillations are very small, and one oscillatory term from the most relevant primary field is sufficient to describe the data.  However, precisely because the oscillations are small, we are unable to reliably extract the scaling dimensions from our numerical fits (though it must be included for reliable extraction of the central charge).  Higher values of the R\'{e}nyi index indeed make it easier to extract exponents from the oscillations; however, this route is impractical since finite size effects become greater for larger $\alpha$.  The benefit of considering periodic boundaries is that it leads to very accurate estimates of the central charge with minimal finite size effects, as shown in Fig. \ref{fig:REE} and Table \ref{tab:central}.  
\begin{table}
\begin{tabular}{cccc} \hline
$N$ & $L$ & $c$ & $c_{\rm CFT}$ \\
\hline
$2$ & $64$ & $0.99(1)$ & $1$   \\
$3$ & $74$ & $2.01(1)$ & $2$   \\
$4$ & $64$ & $2.99(1)$ & $3$   \\
$5$ & $70$ & $3.99(1)$ & $4$   \\
$6$ & $72$ & $5.01(1)$ & $5$   \\
\hline
\end{tabular}
\caption{Best fit central charges corresponding to Fig. \ref{fig:REE}.  Exact values are given by $c_{\rm CFT}=N-1$.  These results are obtained by excluding the first few data points when fitting to the form Eq.~(\ref{eq:GenScalingForm}).}
\label{tab:central}
\end{table} 

\subsection{REE open boundaries, $N<4$} 
\label{subsec:REEopenBCNls4}

\begin{figure}
\centerline{\includegraphics[angle=0,width=1.0\columnwidth]{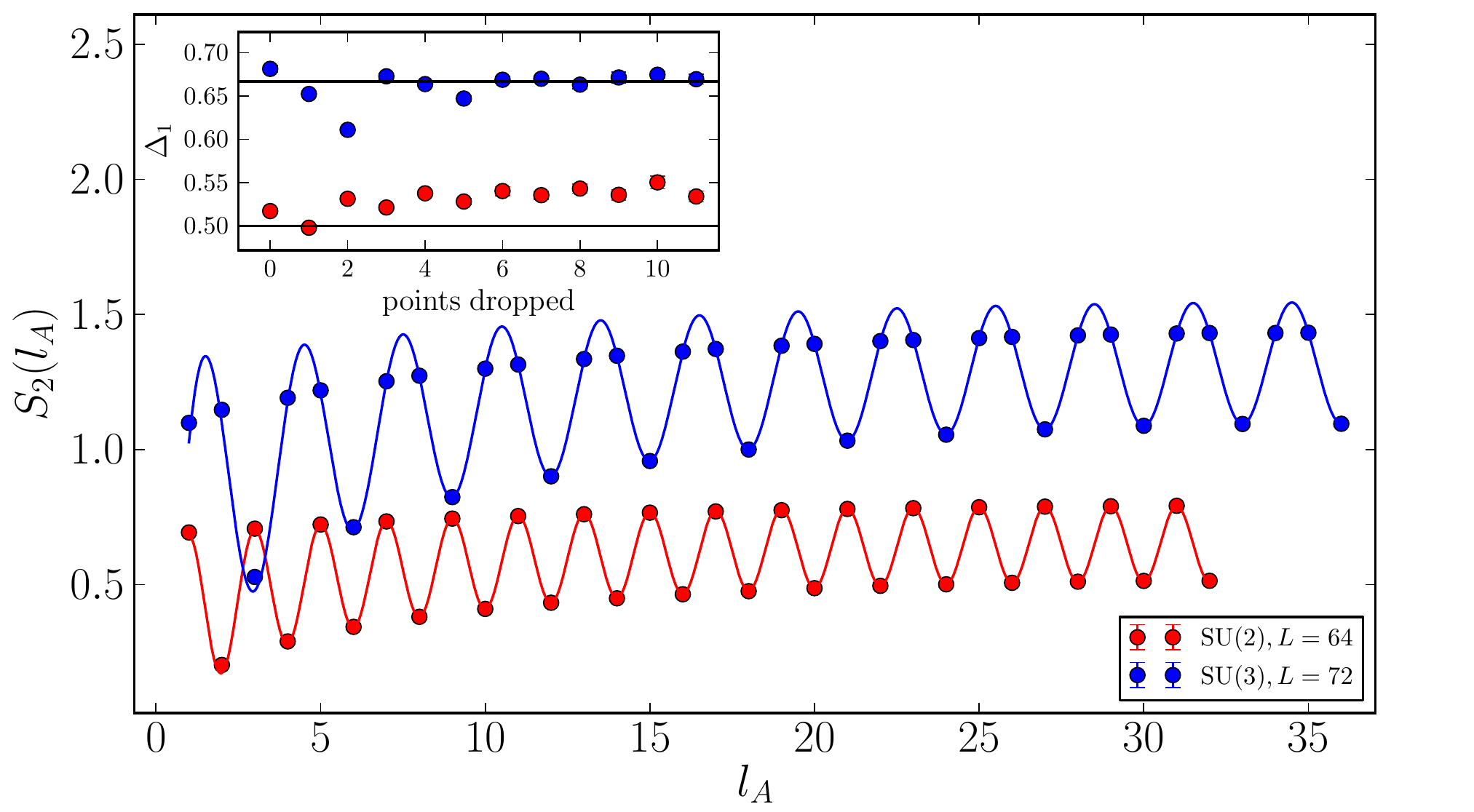}}
\caption{Second R\'{e}nyi entropy with open boundaries for $N=2,3$.  In the inset we plot the scaling dimension ($\Delta_{1}$) as obtained by fitting the QMC data. The solid black lines are the exact values, and the QMC results are plotted as a function of the number of boundary points that are excluded from the fit.}
\label{fig:su2su3Kll}
\end{figure}

In order to efficiently extract the scaling dimensions from the REE, we consider open boundary conditions where oscillations are much more pronounced.  We begin with $N=2,3$ where there is just one distinct scaling dimension.  The second R\'{e}nyi entropy along with the best fit of $\Delta_{1}$ is given in Fig. \ref{fig:su2su3Kll}.  We find that $\Delta_{1}$ in the SU(2) case is not fully converged due to the presence of logarithmic corrections to correlations that have not been accounted for.\cite{Affleck1999:LogCorrections}  Interestingly, this seems to have less of an effect in the SU(3) case where the best fit in the region $l_{A}\gg1$ converges close to the analytical value in the thermodynamic limit (see Inset of Fig.~\ref{fig:su2su3Kll}).  

\begin{figure}
\centerline{\includegraphics[angle=0,width=1.0\columnwidth]{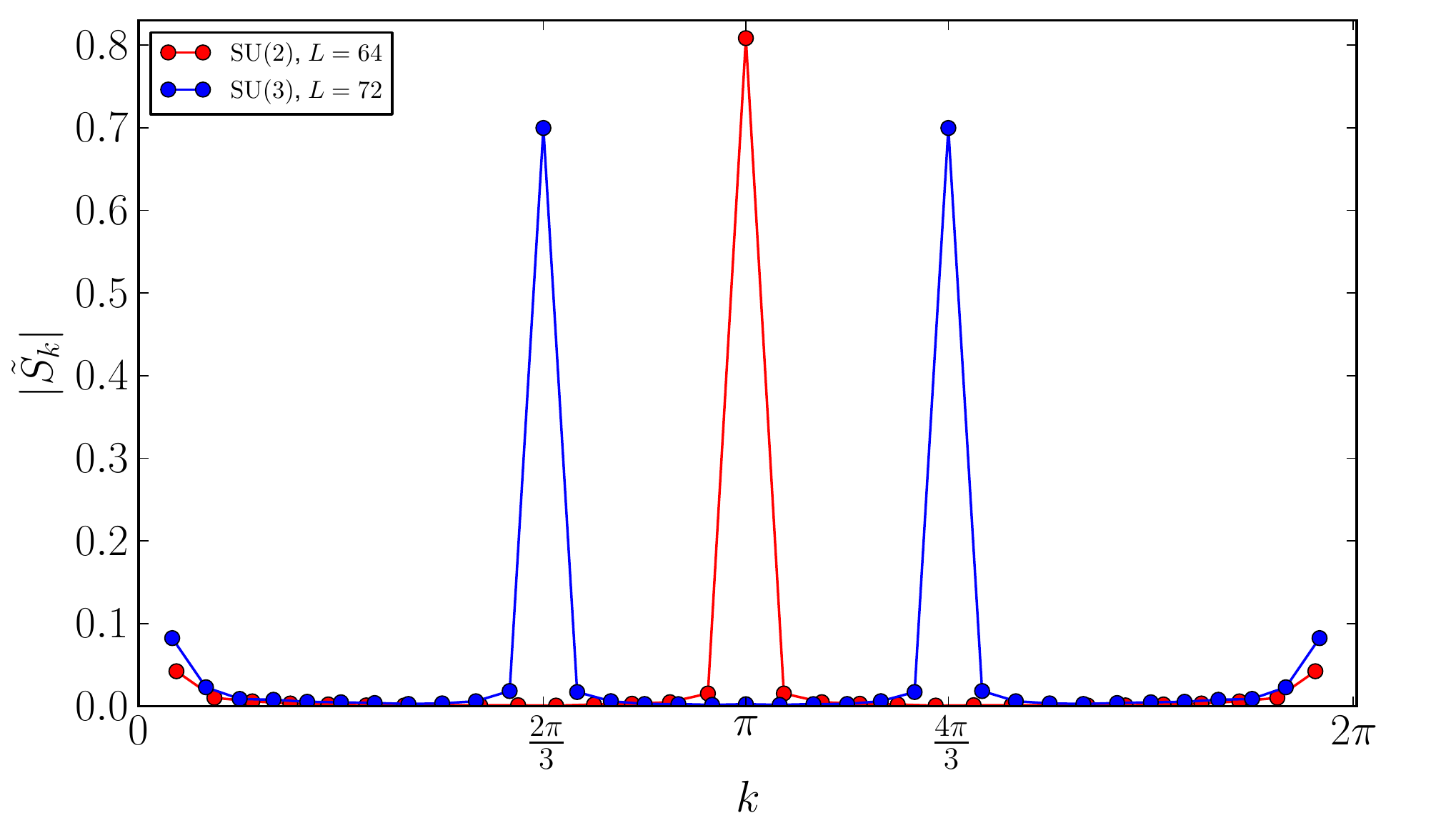}}
\caption{Discrete Fourier transform Eq.~(\ref{eq:EEfourier}) of the REE data appearing in Fig. \ref{fig:su2su3Kll}.  Here we used the fact that $S_{\alpha}(l_{A},L)=S_{\alpha}(L-l_{A},L)$ to reconstruct the REE along the entire chain length, then dropped $L/4$ points from each edge before taking the Fourier transform. Peaks in the Fourier transform appear at integer multiples of 2$k_{F}$.}
\label{fig:EEfouriersu2su3}
\end{figure}

In order to see the qualitative signature of the primary fields, we show in Fig. \ref{fig:EEfouriersu2su3} the discrete Fourier transform of the REE appearing in Fig. \ref{fig:su2su3Kll}.  Before taking the Fourier transform we used the fact that $S_{\alpha}(l_{A},L)=S_{\alpha}(L-l_{A},L)$ to reconstruct the REE along the entire chain length.  We then dropped $L/4$ points from each edge in order to minimize finite size effects coming from the boundary.  The Fourier transform is given by
{\allowdisplaybreaks
\begin{equation}
		\tilde{S}_{k} =\frac{1}{\sqrt{n}}\sum_{j=0}^{n-1}S_{j}e^{-2\pi i k j /n},
\label{eq:EEfourier}
\end{equation}}%
where $S_{j}:j\in\lbrack0,n-1\rbrack$ is the list of entries in $S_{\alpha}(l_{A})$ after the points have been dropped and n is the total number of elements left.  We have dropped the $\alpha$ index from the discrete Fourier transform for ease of notation.


\subsection{REE open boundaries, $N\geq4$} 
\label{subsec:REEopenBCNgeq4}
We now move to $N=4$ which is the first $N$ for which there is a more than one distinct scaling dimension. We hence have to generalize Eq.~(\ref{eq:Sosc}) to a sum of oscillating terms. We use the following form, 
{\allowdisplaybreaks
\begin{equation}
		S^{\mathrm{osc}}_{\alpha}(l_{A}) =\sum_{a=1}^{N-1} f_{\alpha}^{a}\cos(2ak_Fl_{A})\left|\frac{2\eta L}{\pi}\sin(\pi l_{A}/L)\right|^{-\frac{2\Delta_{a}}{\eta\alpha}},
\label{eq:Sosc2}
\end{equation}}%
where we achieve very good fits to our QMC data, again taking the universal scaling function to be a constant. 

\begin{figure}
\centerline{\includegraphics[angle=0,width=1.0\columnwidth]{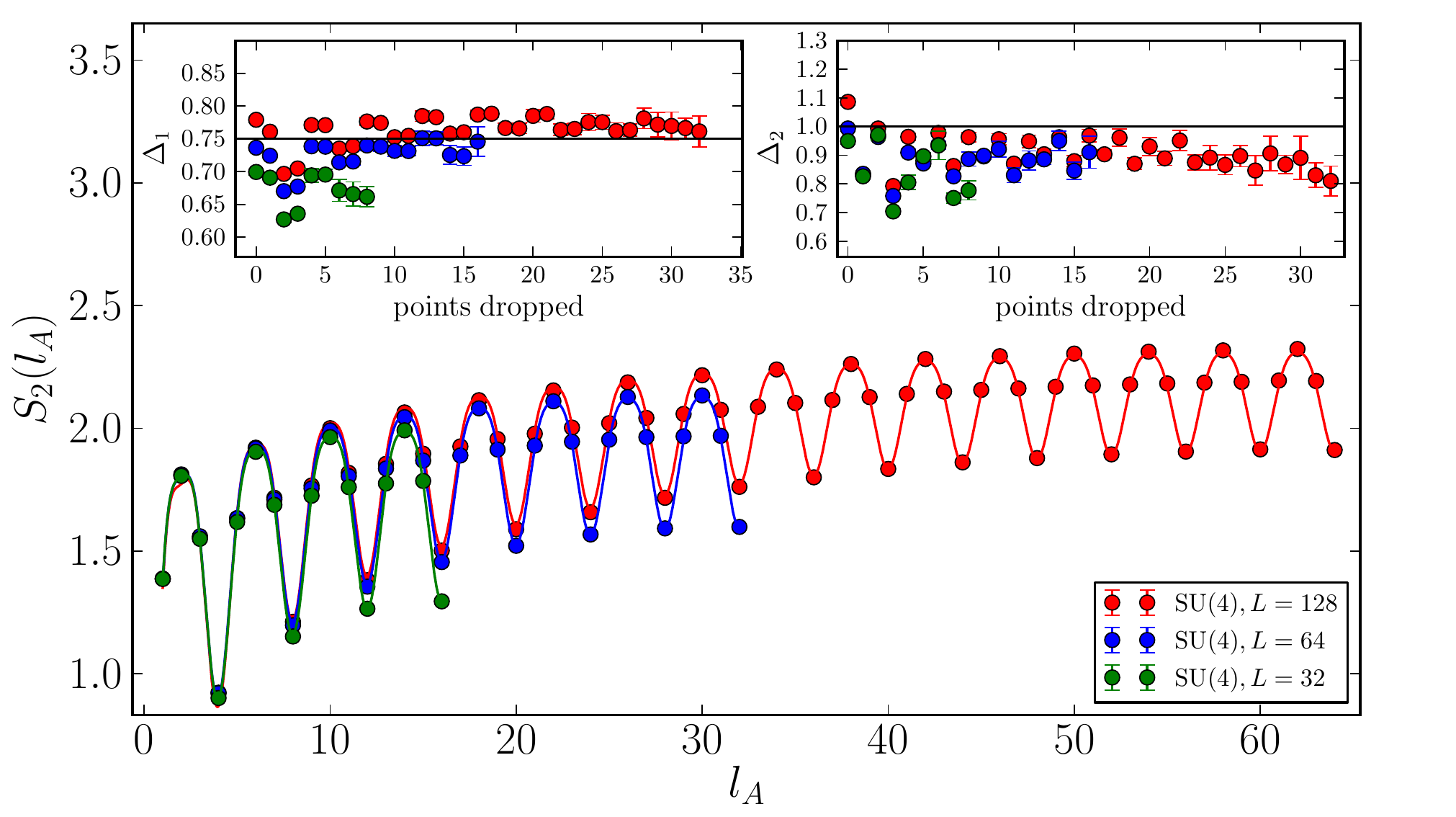}}
\caption{Second R\'{e}nyi entropy with open boundaries for $N=4$ and $L=32,64,128$. The inset is similar to Fig.~\ref{fig:su2su3Kll}, although now we fit two different primary field scaling dimensions. Strong finite size effects are apparent in the extraction of $\Delta_{2}$, the signature of which is much weaker than that of $\Delta_{1}$.  Error bars indicate stochastic error and do not include the systematic error inherent in the fit.}
\label{fig:su4L32L64L128}
\end{figure}

\begin{figure}
\centerline{\includegraphics[angle=0,width=1.0\columnwidth]{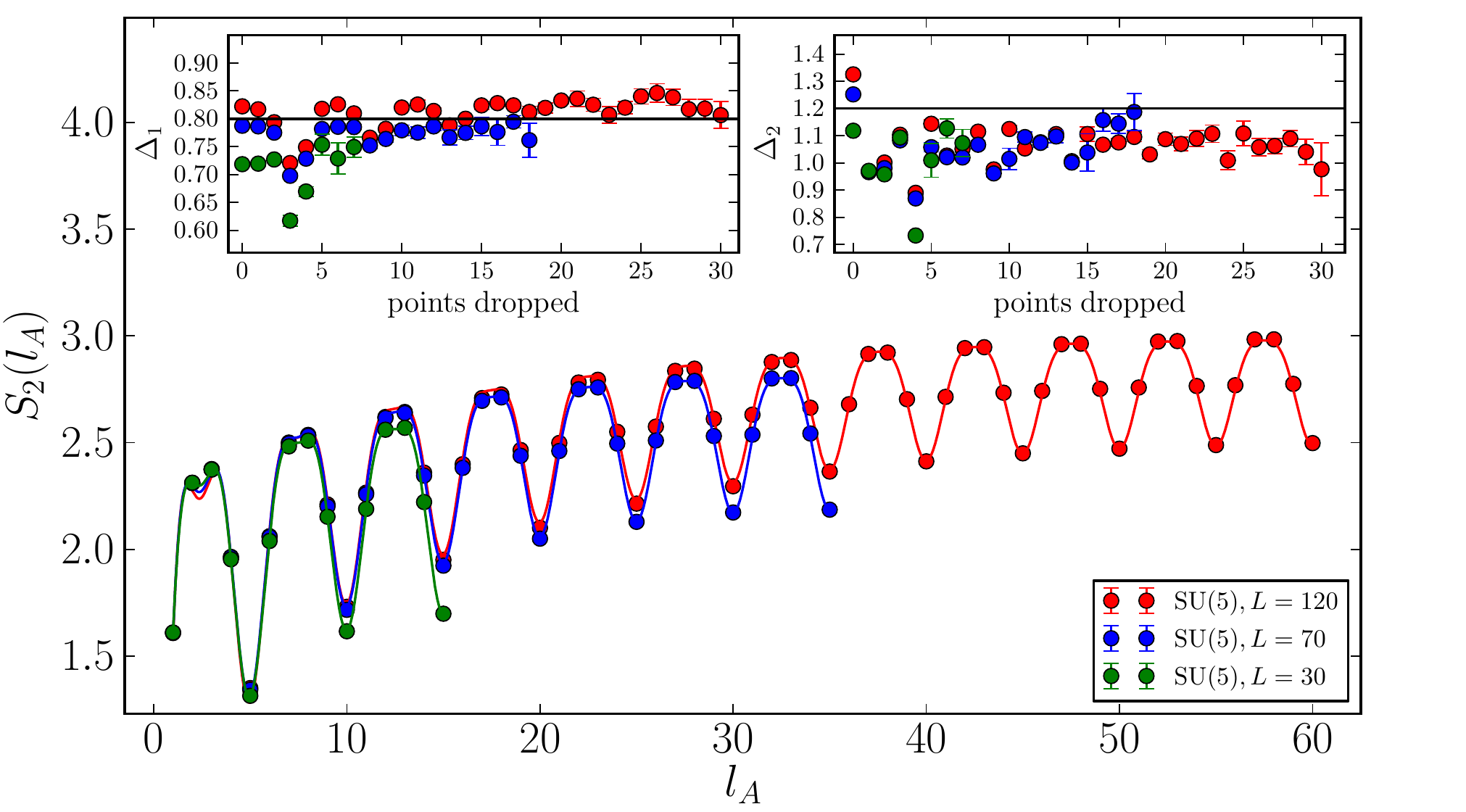}}
\caption{Second R\'{e}nyi entropy with open boundaries for $N=5$ and $L=30,70,120$. This figure is similar to Fig. \ref{fig:su4L32L64L128}.}
\label{fig:su5L30L70L120}
\end{figure}

In Figs. \ref{fig:su4L32L64L128} and \ref{fig:su5L30L70L120} we have fit our data from $N=4$ and $N=5$ with the oscillatory piece Eq.~(\ref{eq:Sosc2}) and use it to extract the two distinct scaling dimensions.  For $N=4,5$ we find it necessary to go to even larger system sizes in order to show convergence of the scaling dimensions to their CFT values.  Strong finite size effects are apparent in the extraction of $\Delta_{2}$, however it is essential to include it as a fit parameter in order to obtain a reasonable estimate of $\Delta_{1}$.


\begin{figure}
\centerline{\includegraphics[angle=0,width=1.0\columnwidth]{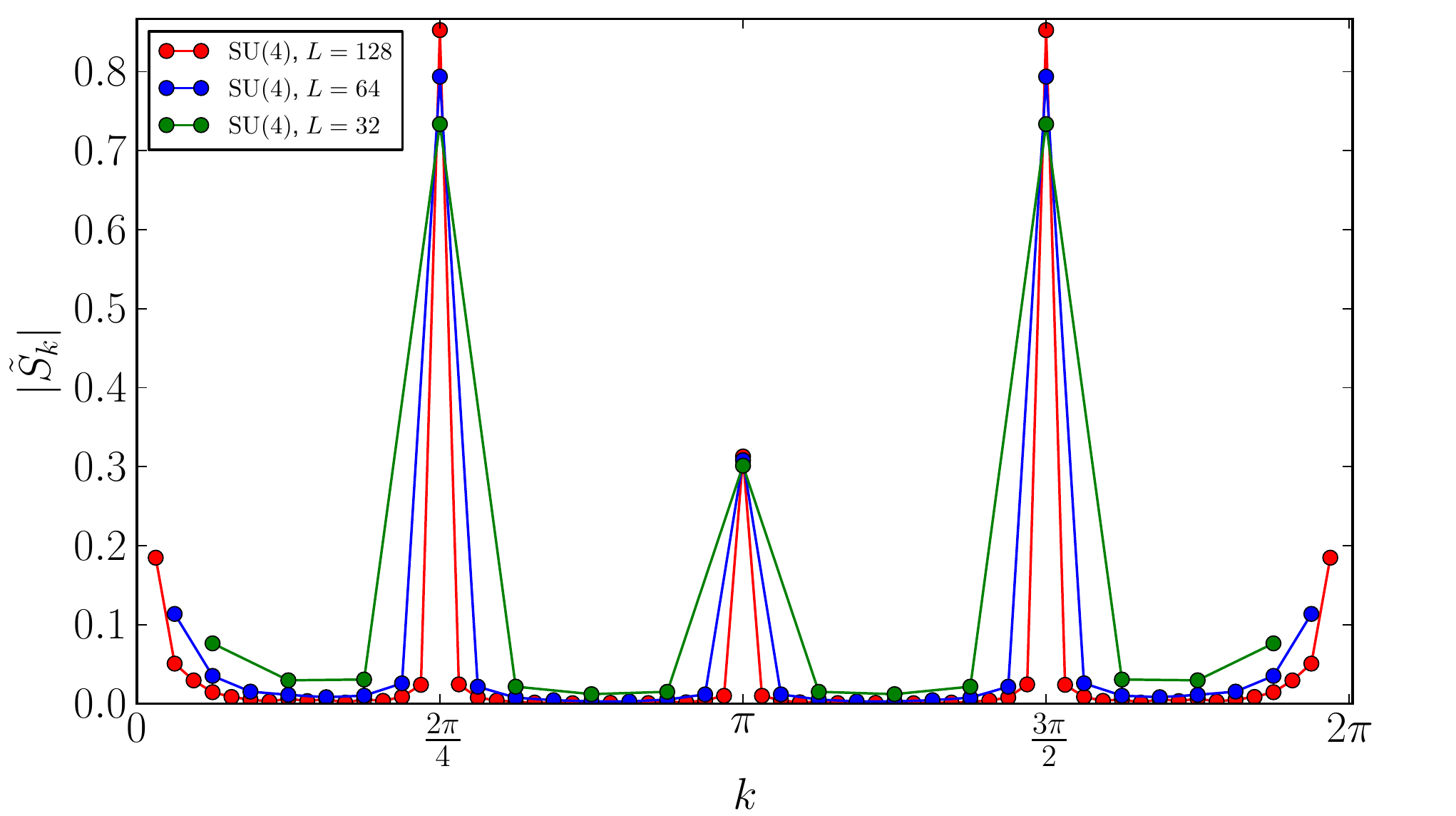}}
\caption{Fourier transform of the second R\'{e}nyi entropy for $N=4$ given in Fig. \ref{fig:su4L32L64L128}.}
\label{fig:EEfouriersu4}
\end{figure}

\begin{figure}
\centerline{\includegraphics[angle=0,width=1.0\columnwidth]{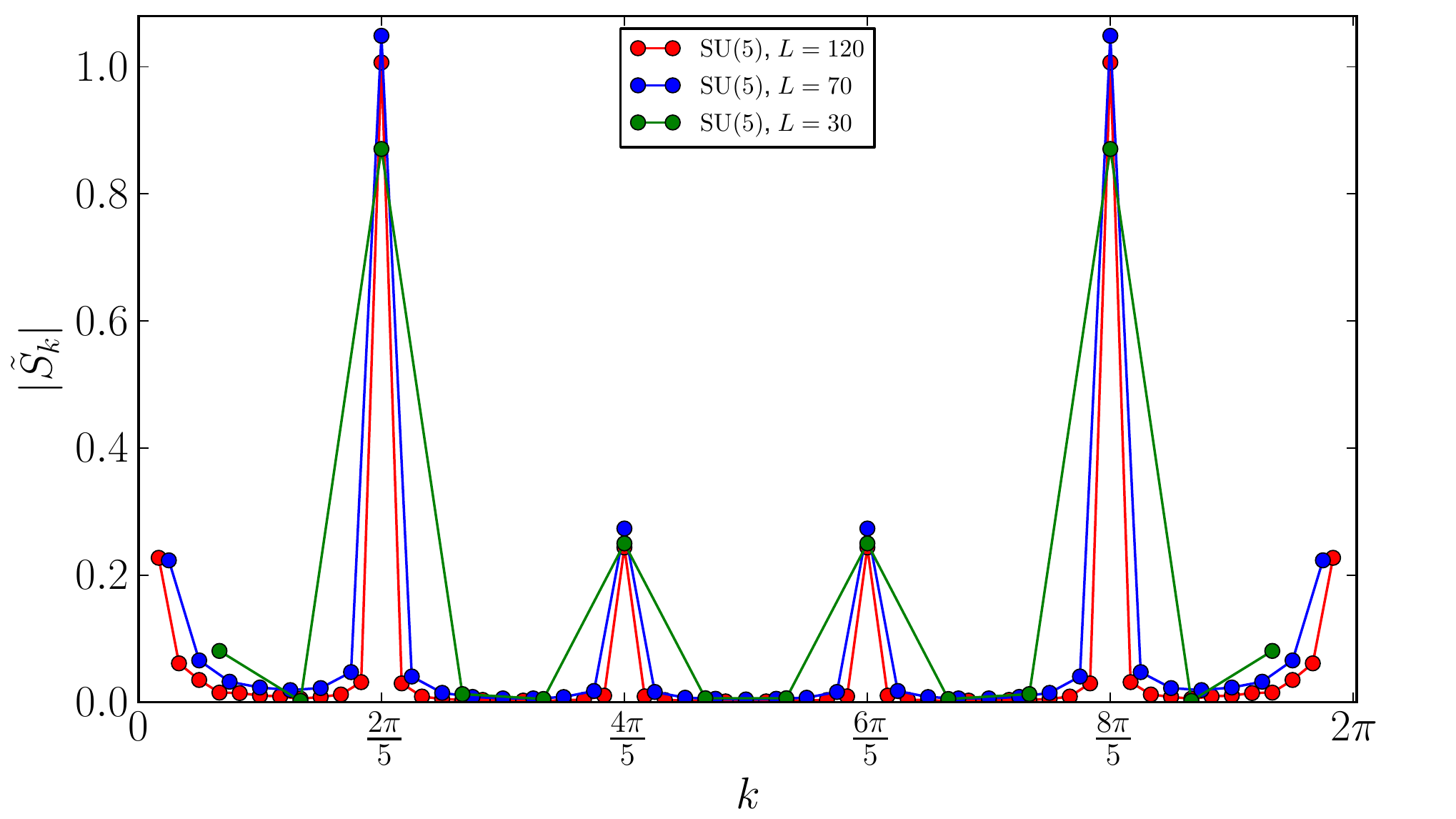}}
\caption{Fourier transform of the second R\'{e}nyi entropy for $N=5$ given in Fig. \ref{fig:su5L30L70L120}.}
\label{fig:EEfouriersu5}
\end{figure}

Though it is difficult to obtain accurate quantitative estimates of $\Delta_{2}$ using this method, clear qualitative signatures of all primary fields are present in the Fourier spectrum of the REE shown in Figs. \ref{fig:EEfouriersu4} and \ref{fig:EEfouriersu5}. 

\subsection{Effect of finite temperature on R\'{e}nyi entropy} 
\label{subsec:thermalRE}
Finally we consider the thermal R\'{e}nyi entropy, which is defined in the same way as Eq.~(\ref{eq:REE}), except that the density matrix is no longer pure (constructed from only the ground state) but rather it is mixed (constructed from a thermal distribution of excited states):
{\allowdisplaybreaks
\begin{equation}
		\rho=\frac{e^{-\beta H}}{Z}.
\label{eq:Rhothermal}
\end{equation}}%
The thermal R\'{e}nyi entropy allows us to extract the central charge at finite temperature which in practice is much less computationally intensive.

\begin{figure}
\centerline{\includegraphics[angle=0,width=1.0\columnwidth]{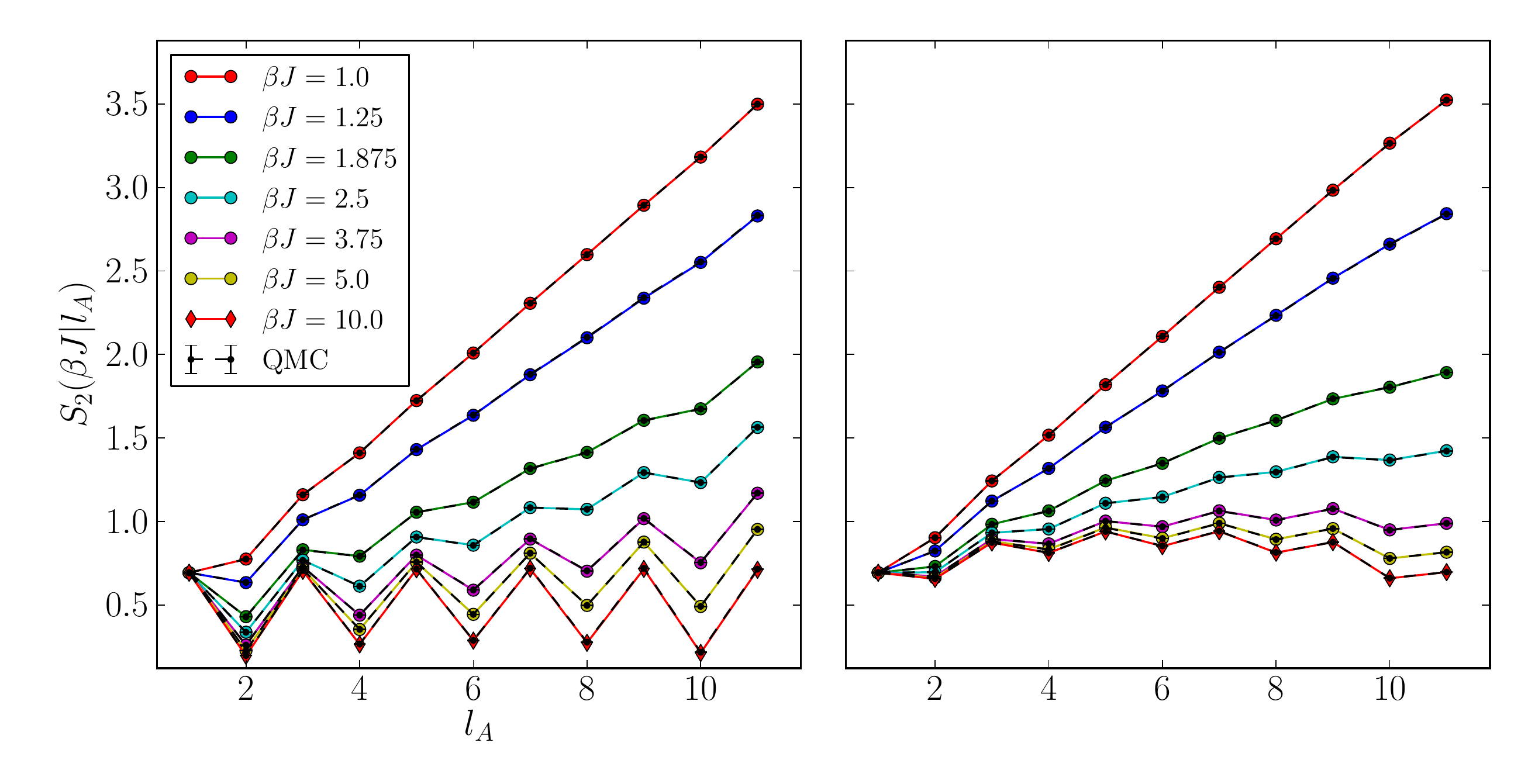}}
\caption{Thermal R\'{e}nyi entropy of a length 12 chain for SU(2) with both open (left) and closed (right) boundaries.  Colored data points are obtained from exact diagonalization, and the black QMC data points are in perfect agreement.   Oscillations still occur at finite temperature, but become drowned out at small enough values of $\beta J$ where the linear scaling regime emerges and open and closed chains take on the same scaling form.}
\label{fig:finiteTemergence}
\end{figure}

Fig. \ref{fig:finiteTemergence} shows the emergence of a linear scaling region in the thermal R\'{e}nyi entropy that now captures the entanglement between subsystems as well as the thermal entropy of subsystem $A$.  Since the entropy goes like the log of the number of states, and in the high temperature limit all of the $N^{l_{A}}$ states are equally probable, we naturally expect a linear scaling region to emerge at finite temperature. 

We use the following scaling form to fit to our thermal R\'{e}nyi entropy data\cite{Calabrese2004:EEandQFT,Korepin2004:thermal}
{\allowdisplaybreaks
\begin{equation}
		S_{2}(\beta |l_{A}) \sim \left (1+\frac{1}{\alpha}  \right ) \frac{\pi c  l_{A}}{12 k_{F} \beta },
\label{eq:Sthermal}
\end{equation}}%
where we fix $k_{F}=\pi/N$.  We tested that this formula gives the correct central charge in the linear scaling regime for different values of $N$ and $\alpha$ and for both open and closed boundary conditions.  Here we only present data for $N=3$ and $\alpha=2$.

\begin{figure}
\centerline{\includegraphics[angle=0,width=1.0\columnwidth]{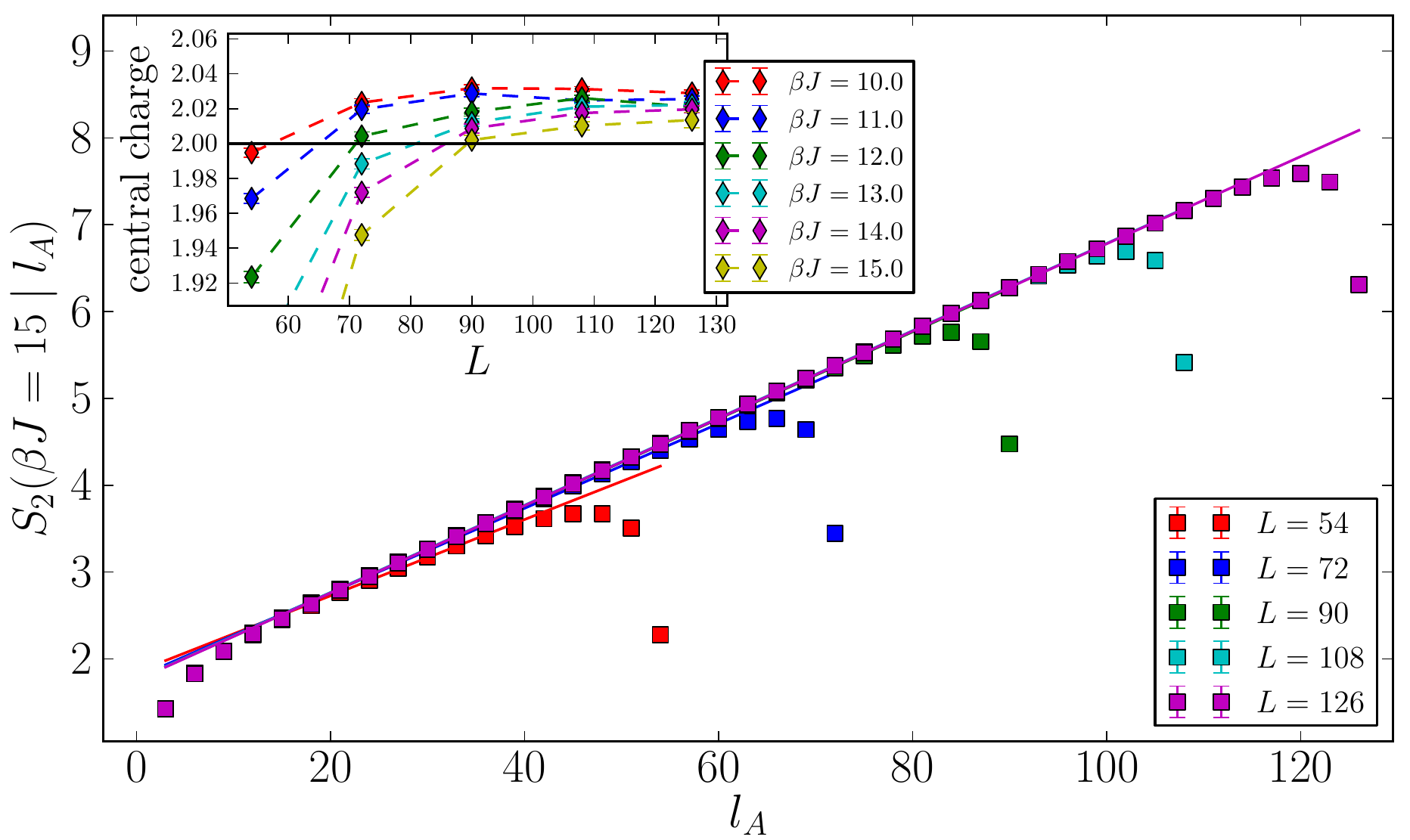}}
\caption{Thermal R\'{e}nyi entropy at $\beta J = 15$ as a function of subsystem size plotted for several lengths on  periodic SU(3) chains.  The inset shows effective central charges extracted using the form Eq.~(\ref{eq:Sthermal}) for several values of total length and $\beta J$.  The fits only include points that are well within the linear scaling regime.}
\label{fig:finiteTsu3}
\end{figure}

In Fig. \ref{fig:finiteTsu3} we use the scaling form Eq.~(\ref{eq:Sthermal}) to extract an effective central charge for $N=3$ chains at different values of $L$ and $\beta J$, which is given in the inset.  We clearly see that the central charge flows to its analytical value in the limit $L \to \infty$ and $\beta J \gg 1$.  In practice, we find the central charge approaches a value slightly higher than its analytical one.  This is due to the fact that oscillations have been neglected by considering only troughs, leading to a linear fit with a slightly larger slope.  
\section{Conclusions} 
\label{sec:conclusions}
In this paper we have investigated the R\'{e}nyi entanglement entropy in the context of critical SU($N$) spin chains which are described by a WZW non-linear sigma model in the thermodynamic limit.  We showed that signatures of all $N-1$ primary fields are present in the oscillations of the entanglement entropy.  We further used the analytical form of the oscillations given by Eq.~(\ref{eq:Sosc2}) to extract the numerical values of the scaling dimensions, which are consistent with the results of CFT.  

We considered both closed and open boundary conditions, where the former proves effective in extracting the central charge, while the latter is more suitable for extracting the scaling dimensions of primary fields.  Finally, we demonstrated universal behavior of the thermal R\'{e}nyi entropy that allows for extraction of the central charge with less computational effort.

These results serve to illustrate the wealth of information contained in the entanglement entropy.  By measuring this quantity alone, one determines all the parameters that make up the continuum description in terms of a CFT.  One could extend this work by considering sub-leading (possibly oscillating) terms in the entanglement entropy for different representations of SU($N$). Such models are described by more general WZW CFTs.  These have been studied numerically in Ref. [\onlinecite{Fuhringer2008:DMRGSUN}] and Ref. [\onlinecite{Rachel2009:SpinonConfinement}], though it would be interesting to see the structure of oscillations that one observes and whether it is possible for scaling dimensions to be extracted via some generalization of Eq.~(\ref{eq:Sosc2}).

Partial financial support was received through NSF DMR-1056536. This work used the Extreme Science and Engineering Discovery Environment (XSEDE), which is supported by National Science Foundation grant number ACI-1053575; in particular, resources were used on the Trestles cluster housed at the San Diego Supercomputing Center (SDSC) allocated under TG-DMR130040. Part of this work was completed while one author (RKK) held
an adjunct faculty position at the TIFR.

\appendix
\section{Stochastic series expansion}\label{app_sse}

In this section, we briefly review the stochastic series expansion (SSE) (see [\onlinecite{sandvik2010:vietri}] for a comprehensive review).  Consider a general spin Hamiltonian which is a sum of bond operators
{\allowdisplaybreaks
\begin{equation}
		H= -\sum_{b}(H^{1}_{b}+H^{2}_{b}-C \mathds{1}),
\label{eq:SSEHamiltonian}
\end{equation}}%
where $H^{1}_{b}$ is diagonal in the spin basis, $H^{2}_{b}$ is off-diagonal and $b$ is the bond index.  Here $C$ is a constant which is adjusted such that the matrix elements of $H^{1}_{b}\geq0$.  Configurations in the SSE come from expanding the partition function,
{\allowdisplaybreaks
\begin{align}
Z=\text{Tr} \{ e^{-\beta H} \} &=\sum_{\alpha} \sum_{n}\frac{\beta^{n}}{n!}\langle\alpha|\{\sum_{b}(H^{1}_{b}+H^{2}_{b})\}^{n}|\alpha\rangle \\
   &=\sum_{\alpha} \sum_{n} \sum_{\{ S_{n} \} }\frac{\beta^{n}}{n!}\langle\alpha|S_{n}|\alpha\rangle,
   \label{eq:SSEExpansion}
\end{align}}%
where $S_{n}$ is a product of $n$ diagonal and off-diagonal bond operators and the sum on $\{ S_{n} \}$ includes all possible combinations of such operators.  We have neglected the constant shift $C$, which should be taken into account when computing the energy.  At this point we should mention that since matrix elements of diagonal operators can always be made positive by the constant shift, minus signs only enter the SSE through off-diagonal operators.  One is then restricted to models where $H^{2}_{b}\geq0$, or models where the number of negative off-diagonals in any configuration is strictly even.  The latter argument will be used in Appendix \ref{app_sampling} to apply the SSE to our antiferromagnetic spin Hamiltonian.

\begin{figure}
\centerline{\includegraphics[angle=0,width=.5\columnwidth]{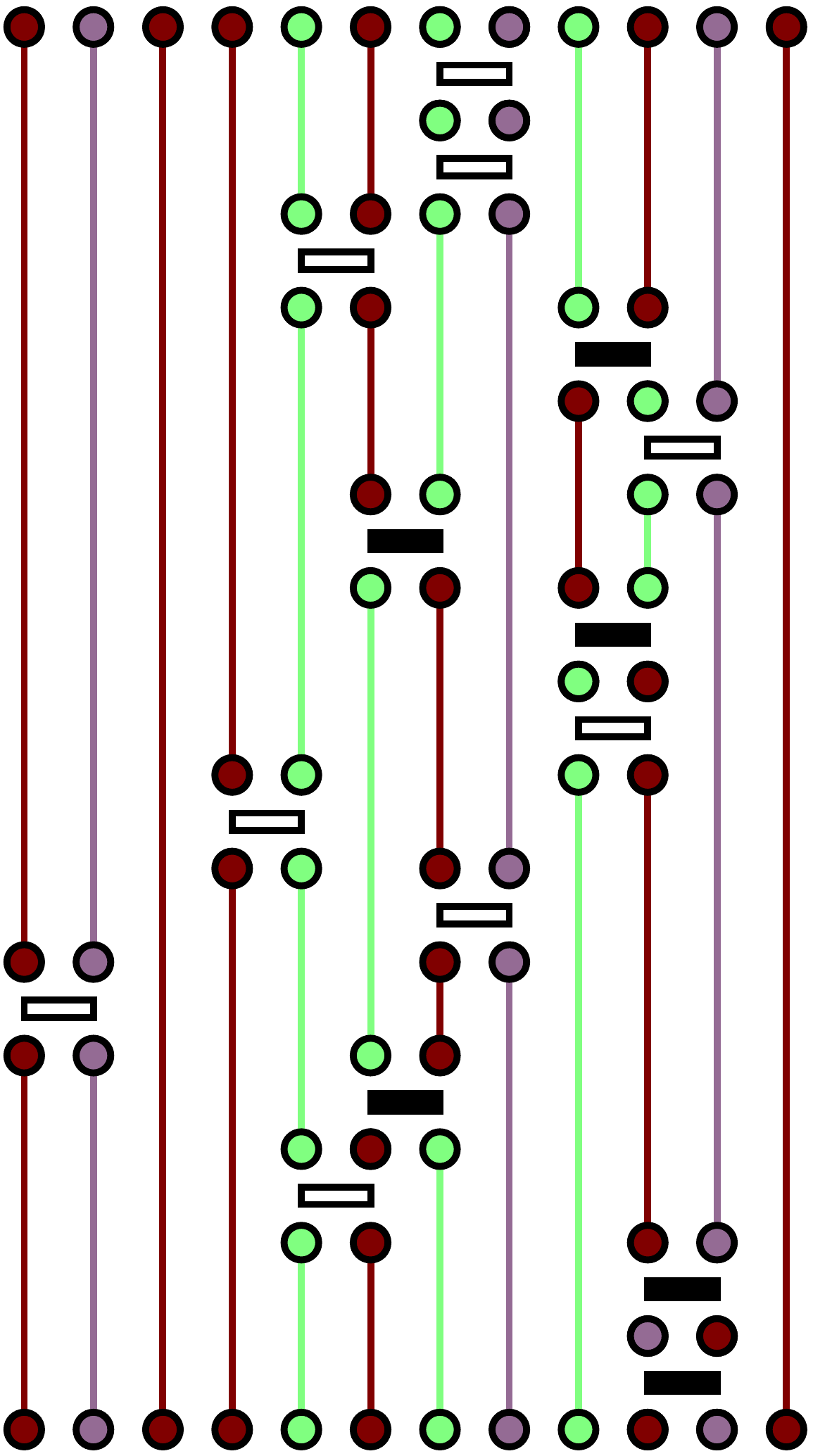}}
\caption{A partition function configuration in the SSE.  Black bars represent off-diagonal operators, and white bars represent diagonal operators.  The operator string satisfies the trace condition, meaning the initial spins at the bottom are the same as the spins at the top.}
\label{fig:Zconfiguration}
\end{figure}

It is useful to introduce a pictorial representation of a configuration in the SSE, which is given in  Fig \ref{fig:Zconfiguration}.  Here we have black (white) bars representing off-diagonal (diagonal) operators in the spin basis, and spin colors are propagated through an operator string that satisfies the trace condition.  The updating scheme for generating new SSE configurations will be given in the context of our SU($N$) model in Appendix \ref{app_sampling}.

\section{QMC sampling procedure}\label{app_sampling}

Previous works~\cite{Frischmuth1999:ThermoSU4,Messio2012:entropyDep} have simulated $H_\Pi$ by Monte-Carlo methods. Below we summarize how we applied the SSE to $H_\Pi$, Eq.~(\ref{eq:PermutationHamiltonian}), and how MC updates of the configurations were carried out. 

In order to avoid minus signs associated with diagonal matrix elements in the SSE (see Appendix \ref{app_sse}), we shift the Hamiltonian by the unit matrix
{\allowdisplaybreaks
\begin{align}
		\mathds{1}=&\sum_{\alpha,\beta} |\alpha\beta\rangle\langle\alpha\beta| \\
		H^{'}_{\Pi} =H_{\Pi}-N_{\mathrm{bond}}\mathds{1}=& \sum_{<ij>}\sum_{\alpha\neq\beta}(|\beta_{i}\alpha_{j}\rangle\langle\alpha_{i}\beta_{j}|\\\nonumber&\qquad\qquad-|\alpha_{i}\beta_{j}\rangle\langle\alpha_{i}\beta_{j}|).
\label{eq:ShiftedPermutationHamiltonian}
\end{align}}%
We thus zero out all matrix elements between same-spin nearest neighbors.  The off-diagonal (spin flip) operators still contribute to minus signs in the SSE, thus any operator string that satisfies the trace and that has an odd number of off-diagonal operators will have a negative weight.  On a chain with open boundary conditions, all configurations that satisfy the trace have an even number of off-diagonal operators.  This can be seen from the fact that there are no matrix elements between same-spin nearest neighbors so that each spin must return to its original position after propagation through the operator string, which requires an even number of permutations.  

With periodic boundary conditions it is possible for like-colored spins to trade places, which in general could be done by an odd number of off-diagonal matrix elements.  However, it can be shown that if a configuration has a fixed parity, i.e., the total number of each different type of color is either all odd, or all even, then such negative weight windings are impossible.  This is precisely what happens in the ground state of chains where the total length is evenly divisible by $N$, because the ground state contains $L/N$ color singlets.  Thus we will sample all configurations with the correct (positive) weight if we remain at low enough temperature to be in the ground state subspace and if we consider chain lengths that are integer multiples of $N$.  In fact we can go further.  After discussing the loop structure in the SSE, we will argue that we avoid a sign problem even at finite temperature in the periodic case.  

\begin{figure}
\centerline{\includegraphics[angle=0,width=.7\columnwidth]{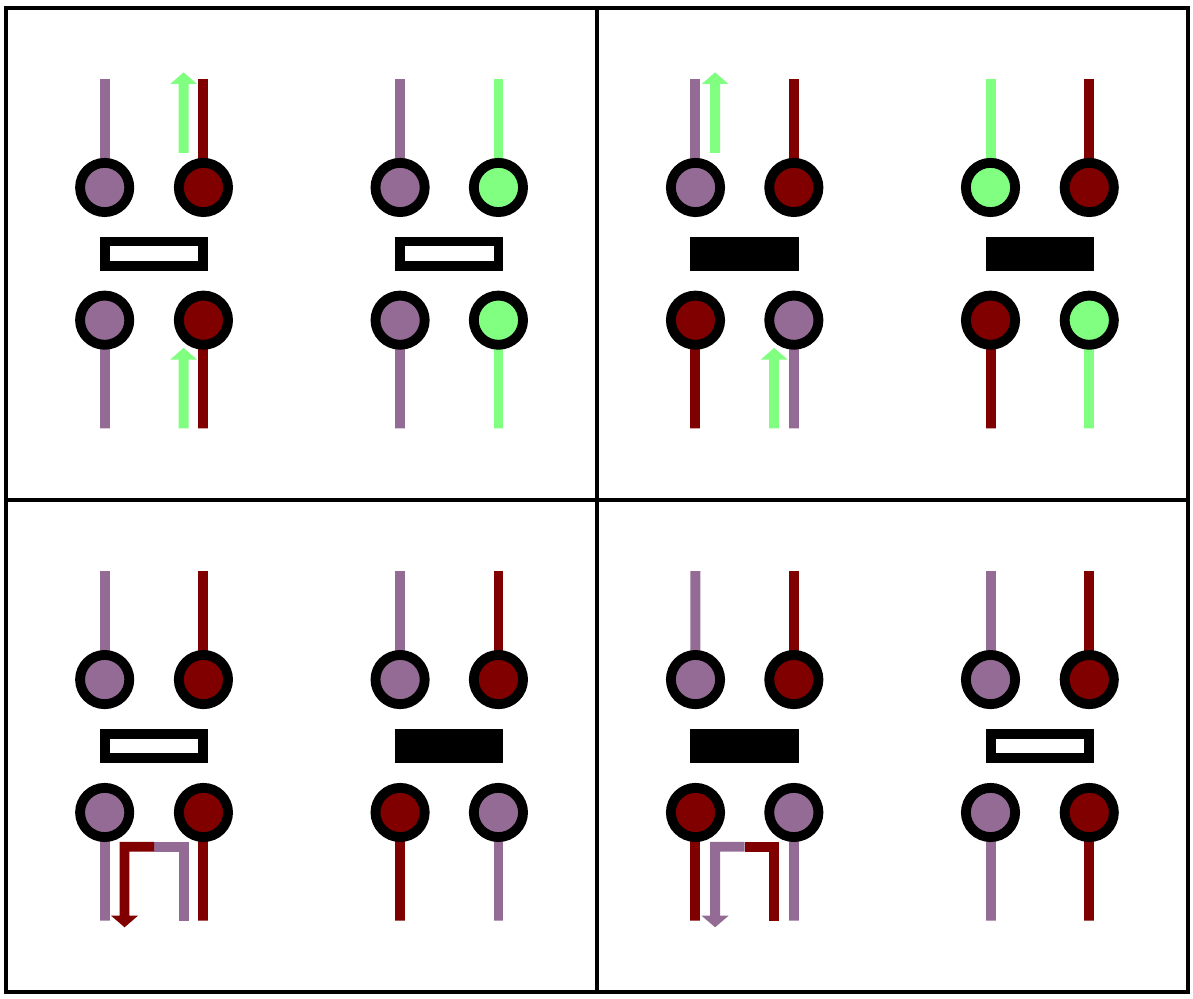}}
\caption{Possible vertices of the Hamiltonian and different loop updating moves.  The left side of each box shows the vertex before the update, along with path and color of the loop (arrows).  The right side of each box shows the vertex after it has been updated.  The upper boxes show the ``continue straight"  and ``switch and continue" moves where the incoming color will not generate a zero matrix element.  The lower boxes are both ``switch and reverse" moves that are required to avoid generating zero matrix elements.}
\label{fig:vertices}
\end{figure}

At this point we will review the Monte Carlo sampling procedure, highlighting the novel loop structure that allows for deterministic sampling of configurations.  Monte Carlo sampling occurs in four stages, the first being a diagonal update, where diagonal operators are inserted or removed from the operator string with a probability given by the ratio of weights in the SSE.  Next, linked lines are constructed that establish a loop structure to the configuration.  A random color and starting position is then chosen, and one follows along the loop changing colors according to the starting color until the loop closes.  Painting loops causes the conversion between diagonal and off-diagonal operators, and when the loop closes we generate a new configuration contained in the SSE of the partition function.  Finally, one makes measurements on the newly generated configuration; in the case of the REE, checking if a transition can be made between different partition function ensembles (see Appendix \ref{app_replica}).  Loop moves are constructed so that matrix elements with zero weight are never generated.  In our case, the zero matrix elements are between nearest neighbors with the same color.  Figure \ref{fig:vertices} shows the possible vertex updating moves, or equivalently, how a loop update can change matrix elements within the SSE of the partition function.
  
The upper boxes show updates where the incoming color poses no threat of a zero matrix element.  The lower boxes require a switch and reverse loop move along with a loop color switch so as to avoid creating a zero matrix element.  Once a random starting color and starting position for the loop is chosen, then the path of the loop as well as the sequence of color changes are uniquely determined by these rules.

As we increase the value of $\beta J$, the operator string grows in length (vertical direction of Fig \ref{fig:Zconfiguration}), and the loops increase their spatial (horizontal) extent.  Conversely, loops at high temperature become localized in the horizontal spatial direction, and are less likely to wind around in the horizontal direction.  It is precisely for this reason that the algorithm does not suffer from a sign problem in the case of periodic boundaries at finite temperature.  Negative weight configurations (though possible), have a negligibly small weight.  As the temperature is lowered and the loops begin to proliferate, we begin to enter the total spin zero sector of the configuration space (the same number of every color is present) and negative winding configurations cannot occur.  This is why it is crucial that we have chain lengths which are integer multiples $N$, since free spins (not paired in a singlet) are free to wander and create negative weight windings.

\section{REE and the replica trick}\label{app_replica}
\begin{figure}[h]
\centering
\begin{tabular}{m{.13\textwidth} m{.07\textwidth} m{.13\textwidth}} 
\includegraphics[width=.13\textwidth]{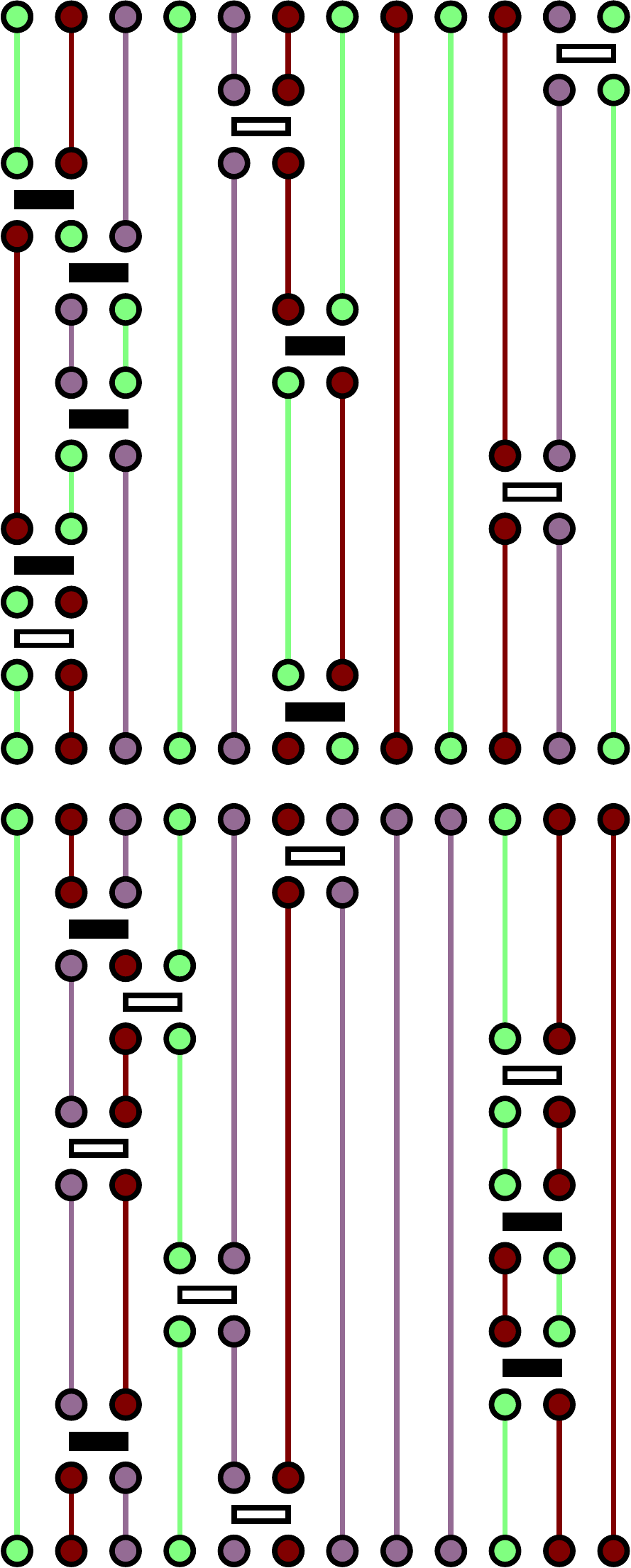} & 
\includegraphics[width=.07\textwidth,height=.03\textheight]{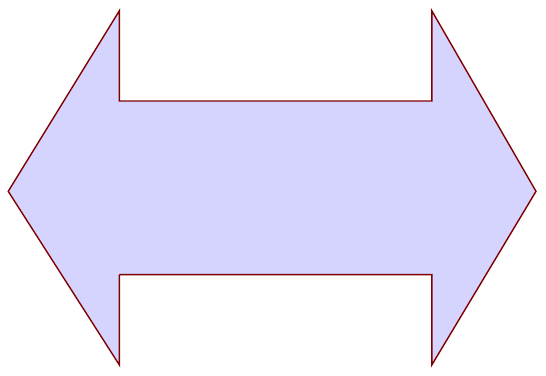} &
\includegraphics[width=.13\textwidth]{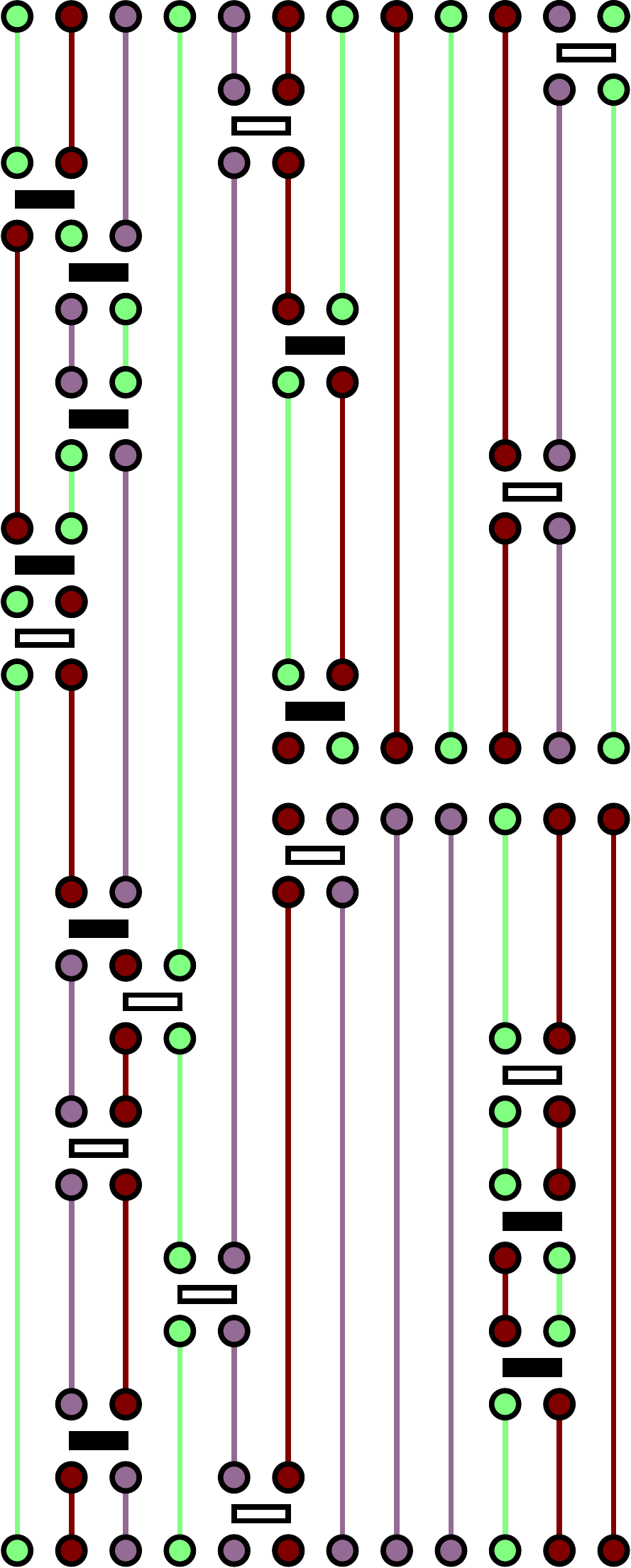} 
\end{tabular}
\caption{Configuration of $Z^{2}$ (left) and $Z^{(2)}_{A}$ (right).  A transition from left to right can be made when the $A$ spins in both copies match each other.  Transitions from right to left are made when the $A$ spins in the middle match the $A$ spins in the outer trace. }
\label{fig:ExtendedEnsemble}
\end{figure}

We will now review the recipe for measuring the R\'{e}nyi entanglement entropy within the context of the SSE.  The two basic ingredients are the replica trick, \cite{Melko2010:FiniteTRenyi,Hastings2010:MeasuringRenyiQMC} and an extended ensemble QMC approach introduced in [\onlinecite{Humeniuk2012:ExtendedEnsemble}].  The R\'{e}nyi entanglement entropy is given by
{\allowdisplaybreaks
\begin{equation}
		S_{\alpha}(\rho_{A})= \frac{1}{1-\alpha}\log(\text{Tr}\{ \rho^{\alpha}_{A} \}),
\label{eq:RenyiEntropy1}
\end{equation}}%
where
{\allowdisplaybreaks
\begin{equation}
		\rho_{A}=\frac{1}{Z}\text{Tr}_{B}\{ e^{-\beta H} \}
\label{eq:ReducedDensityMatrix}
\end{equation}}%
is the reduced density matrix, given by tracing over the basis states in the $B$ subsystem.  When considering the entanglement between $A$ and $B$ at zero temperature, we need to have $\beta$ sufficiently large so as to project out only the ground state contribution to Eq.~(\ref{eq:ReducedDensityMatrix}).  We can express the entanglement entropy in terms of a ratio of partition functions\cite{Humeniuk2012:ExtendedEnsemble}
\begin{equation}
S_{\alpha}(\rho_{A})= \frac{1}{1-\alpha}\log\frac{Z^{(\alpha)}_{A}}{Z^{\alpha}},
\label{eq:RenyiEntropy2}
\end{equation}
where $Z^{\alpha}$ consists of $\alpha$ copies of the regular partition function, and $Z^{(\alpha)}_{A}$ has a modified trace condition in the $A$ sub-region, which extends over all copies.  An example of each type of configuration is given in Fig. \ref{fig:ExtendedEnsemble} for the case $\alpha=2$.  

To compute the Monte Carlo average of the ratio of partition functions, one preforms importance sampling of the extended ensemble of configurations contained in $Z^{\alpha} \cup Z^{(\alpha)}_{A}$.  Transitions between ensembles can be made when the spin matching condition is satisfied in the $A$ sub-region.  In this way, the ratio is given by the number of Monte Carlo steps preformed in $Z^{(\alpha)}_{A}$, divided by the number of Monte Carlo steps preformed in $Z^{\alpha}$
{\allowdisplaybreaks
\begin{equation}
\Bigg\langle\frac{Z^{(\alpha)}_{A}}{Z^{\alpha}}\Bigg\rangle=\Bigg\langle\frac{MC_{step}(Z^{(\alpha)}_{A})}{MC_{step}(Z^{\alpha})}\Bigg\rangle_{MC}	.
\label{eq:RatioAverage}
\end{equation}}Sampling the extended ensemble in the SSE is particularly straightforward, since transitions between ensembles occurs with probability 1 whenever the spin matching condition is satisfied in the $A$ sub-region.  This is due to the fact that the two configurations have identical weights (number of bond operators), though this idea can be extended to configurations that have continuous degrees of freedom.\cite{Humeniuk2012:ExtendedEnsemble}  In practice, the $A$ spin matching condition becomes increasingly difficult to satisfy as the $A$ sub-region grows, so we employ an increment trick used in both [\onlinecite{Hastings2010:MeasuringRenyiQMC}] and [\onlinecite{Humeniuk2012:ExtendedEnsemble}].  
We note that the modified partition function $Z^{(2)}_{A}$ locally breaks translational invariance at a point joining the $A$ and $B$ subsystems.\cite{Cardy2010:Unusual}  This leads to oscillations in the R\'{e}nyi entanglement entropy under periodic boundary conditions.  The von Neumann entanglement entropy lacks these oscillations under periodic boundary conditions as it is free from this defect.  Both the R\'{e}nyi and von Neumann entanglement entropies can have oscillatory terms under open boundaries due to the breaking of translational invariance at the edges of the chain.

\bibliographystyle{apsrev}
\bibliography{EE_suN_spins_paperFINAL}

\end{document}